# Enhancing Privacy for Biometric Identification Cards

Paul BĂLĂNOIU
Academy of Economic Studies, Bucharest, Romania
paul@balanoiu.com

*Most developed countries have started the implementation of biometric electronic identification cards, especially passports. The European Union and the United States of America struggle to introduce and standardize these electronic documents. Due to the personal nature of the biometric elements used for the generation of these cards, privacy issues were raised on both sides of the Atlantic Ocean, leading to civilian protests and concerns. The lack of transparency from the public authorities responsible with the implementation of such identification systems, and the poor technological approaches chosen by these authorities, are the main reasons for the negative popularity of the new identification methods. The following article shows an approach that provides all the benefits of modern technological advances in the fields of biometrics and cryptography, without sacrificing the privacy of those that will be the beneficiaries of the new system.*

# Introduction

The 2003 European Council of Thessaloniki confirmed the need in the European Union for a coherent approach on biometric data for documents, both for European Union citizens and third country nationals. The main purpose for enhanced and harmonized security standards was the protection against falsification and fraudulent use. The introduction of biometric identifiers was found necessary in order to establish a reliable link between the genuine holder and the document he posses [3].

The United States is also deploying a similar system, through the Department of Home Security and the State Department. This also led to the promulgation by the International Civil Aviation Organization (ICAO) of a standard for e-passports [1].

An important concern is the protection of individuals with regard to the personal data to be processed in the context of passports and other travel documents. No unneeded information should be stored in such documents, and the information shouldn't be available to more persons than necessary.

The information to be associated with such documents includes a facial image on the storage medium, and fingerprint information stored in electronic format with emphasis on the integrity, the authenticity and the confidentiality of the data.

The biometric features stored in these documents should only be used for verifying the following:
• the authenticity of the document
• the identity of the holder

Several member states have already planned the implementation of a central database for storing the biometric data of the passport. Although it is possible to implement only a verification procedure of biometric data using a centralized database, this approach presents additional risks regarding the protection of personal data, such as the development of further purposes not foreseen in the regulation [4].

The right way to implement such a passport would be to use only the decentralized storage in the chip of the passport, and the rest of the article will show how such an approach can be implemented with the currently available technical solutions.

**Public key infrastructure**
The concept of public-key cryptography was introduced by Whitfield Diffie and Martin Edward Hellman in their 1976 paper "New Directions in Cryptography". The public-key (or asymmetric) cryptography uses a pair of mathematically related keys to perform the encryption and decryption operations. One of these paired keys is only known to its owner (the private key) while the other is publicly known (the public key). The introduction of the public key cryptography was a quantum leap in the security field because it offers a practical solution to multiple problems: data and party authentication, privacy without a shared secret and key distribution [7].

A year later, Ron Rivest, Adi Shamir, and Leonard Adleman at the Massachusetts Institute of Technology publicly described the algorithm that had became the de-facto standard for public key cryptography in the last thirty years, RSA (named after the initials of their surnames) [9].

The two main applications of public key cryptography are data confidentiality and data authentication.

For data confidentiality, a message encrypted using the recipient's public key can only be decrypted by the recipient, using the paired private key (see figure 1) [6].

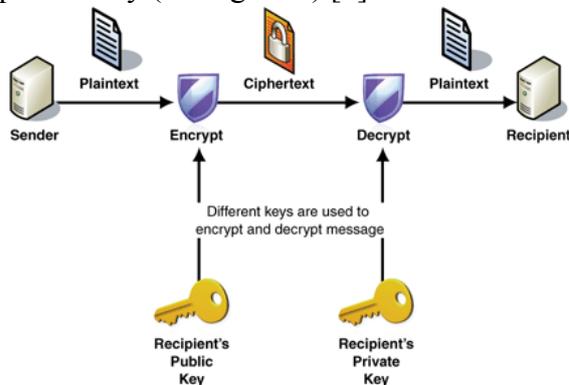

**Fig. 1.** Data confidentiality

For data authentication, a message signed with the recipient's private key can be openly checked by anyone possessing the public key and validated; data tampering is virtually impossible without breaking the validation (see figure 2) [6].

Both applications are of interest for electronic identification documents, as the article will show.

Today, the public key cryptography has become an essential security mechanism for any open and popular system. Even for devices with memory, power, and computational resource constraints, for which the conventional algorithms used in the public key cryptography are not well suited, new methods and algorithms have been developed, such as the Elliptic Curve Cryptography [10].

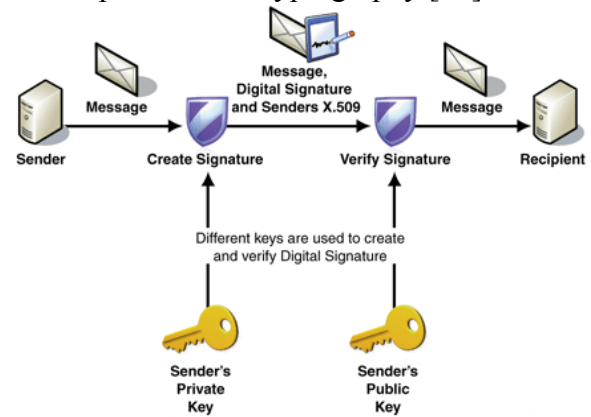

**Fig. 2.** Data authentication

Both applications are of interest for electronic identification documents, as the article will show.

Today, the public key cryptography has become an essential security mechanism for any open and popular system. Even for devices with memory, power, and computational resource constraints, for which the conventional algorithms used in the public key cryptography are not well suited, new methods and algorithms have been developed, such as the Elliptic Curve Cryptography [10].

**Identity Certificates**

Public key certificates (identity certificates) are electronic documents that bind together a public key and an identity. Their main purpose is to verify that a public key belongs to an individual, identified by the information stored in the certificate. That information can consist of the name, the address, unique identification numbers, but can also be extended to include additional data, such as the facial image required by the electronic passport regulations.

In order to verify the information stored in the certificate, a validation mechanism is required. This validation mechanism is the Public Key Infrastructure (PKI) and consists in a set of hardware, software, people, policies and procedures needed to create, manage, store, distribute and revoke digital certificates [10]. It binds the identities and public keys from certificates by means of a certificate authority. The binding is usually performed using the same public key cryptography methods described above, and digitally signing the certificates using the private key of the certificate authority. The process of obtaining the certificate associated with a private key can be seen in figure 3 [6].

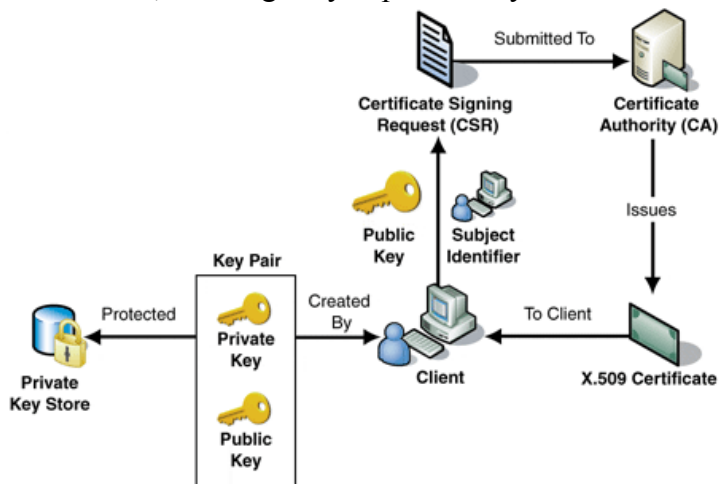

**Fig. 3.** Public Key Infrastructure

The certificate authority that issued a certificate can also revoke it, if the integrity of the certificate has somehow been compromised. This can occur in one of the following circumstances:
- the certificate is no longer required, and the subject will no use it anymore: for example, the person no longer needs the passport
- the identity corresponding to the certificate is removed: for example, the person dies and this makes the passport no longer useful
- the subject of the certificate has breached the trust of the certificate authority that issued the certificate: for example, the identification data associated with the certificate was intentionally misrepresented to the certificate authority during the process of verifying the identity of the subject
- the certificate was used for illicit purposes: for example, the passport is sold and used by other persons, or the international policies are not followed by the owner of the passport
- the private key is compromised, due to theft or improper storage: for example, the passport is stolen from the subject and the risk exists of it being used by other people to identify as the original owner

The list above is by no means exhaustive; other reasons for certification revocation can be added, based on the certificate authority policies.

There are alternatives to the public key infrastructure, such as the web of trust, which involves self-signed certificates and attestations by third-parties of those certificates, or simple public key infrastructure, which doesn't use any notion of trust, the verifier being also the issuer. Neither of these is of interest for electronic passports, since the primary requirement is the authentication of the document by a public agency, so the assumed model for the rest of the article will be the public key infrastructure.

The public agencies will act as certificate authorities, providing all certificate related activities mentioned above, from receiving a certificate request from the person needing a new identification document and issuing the valid certificate, to the management of certificate revocation lists according to public policies.

**Smart cards**

Smart cards are cards that incorporate within their thickness one or more integrated circuits (see figure 4). They are also called chip cards or integrated circuit cards [5].

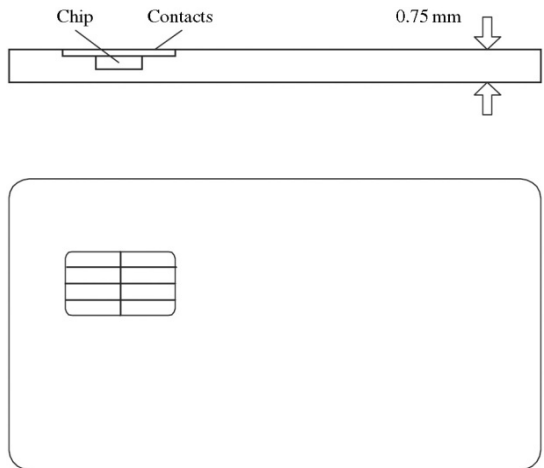

**Fig. 4.** Typical smart card

The most commonly used format for smart cards is defined in the ISO/IEC 7810:2003 international standard that defines four formats of identification cards: ID-1, ID-2, ID-3 and ID-000, especially the ID-1 (banking cards) and ID-000 (SIM cards) [5].

Additional international standards that refer to smart cards:
- ISO 7813 that defines additional characteristics of ID-1 identification cards
- ISO 7811 that defines techniques for recording data on ID-1 identification cards
- ISO 7816 that defines ID-1 identification cards with an embedded chip (smartcard) and contact surfaces
- ISO 14443 defines identification cards with an embedded chip and a magnetic loop antenna

The most appropriate smart cards for using in the identification process are those that include public key cryptography functionalities. Such a smart card can:
- generate a private-public key pair;
- store the private key in an inaccessible store (the key cannot be retrieved from the smart card, it can only be used internally in performing cryptographic functions)
- retrieving the public key for publication
- sign data using the private key
- decrypt data encrypted with the public key
- generate certificate requests containing the public key
- store certificate responses from certification authorities
- besides asymmetric cryptographic functionalities they can usually perform symmetric encryption and decryption
- biometric enrolment

The last function will be further detailed, since it represents one of the most important requirements of the electronic passport.

**Biometrics**
In most scenarios, a smart card is associated with a person. It is very important to be able to ensure that the person using the smart card is the person that the smart card was initially associated with. For some applications, when the smart card main function is to identify the person, like in access control applications or passport applications, this is the main purpose of the smart card itself. In other cases the smart card is used to allow the access to data stored about that person. In both scenarios, it is mandatory that the system ensures the pairing between the person and the smart card.

One of the advantages of a smart card based system is that it doesn't usually need to recognize the person. The common approach is to verify the person identity, to determine if the person is who he or she claims to be, which is much easier and faster than recognition, which involves a great deal of computing time and database searches [5].

The most commonly used human characteristic used for biometric matching is the fingerprint. The rest of the article will refer to fingerprint authentication, unless otherwise stated.

Biometric matching is a complex process, consisting is the following steps [2]:
a) Data acquisition (an image scan of the fingerprint, received from a fingerprint scanner)
b) Data enhancement (edge detection algorithms, resulting a cleaned up version of the scanned image)
c) Feature detection (patterns and minutia identification in the cleaned image)

A visual representation of the feature extraction process can be seen in figure 5.

There are three approaches to adding biometric information to smart card systems [8]:
- Template-On-Card (TOC) – the template is stored on the smart card, but all the biometric

procedures are performed outside the smartcard, at the reader side; these include data acquisition, feature extraction and matching; during this process the reader requests the identifying template from the smart card and matches it with the scanned template; this approach can be applied to any biometric information, not only fingerprints.

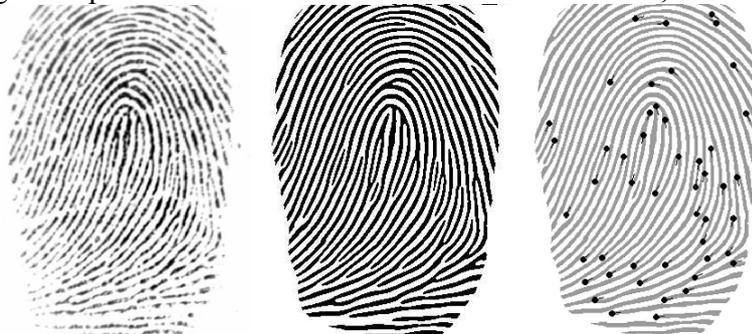

**Fig. 5.** Feature extraction process

- Match-On-Card (MOC) – the original biometric information is stored on the smart card; some of the biometric procedures are performed on the reader side, like data acquisition and feature extraction; this way a new template is constructed for the scanned information; the new template is then sent during for validation to the smart card, which performs a matching check with the internally stored template; the final decision is computed by the smart card itself; the template stored on the smart card is not retrievable, just like the private key used; this approach can also be applied to any biometric information, not only to fingerprints
- System-On-Card (SOC) – the smart card incorporates the original template, but also the entire reader, including the biometric sensor and biometric processor; all validation procedures are performed on the smart card itself: data acquisition, feature extraction, template generation and matching); this approach is applicable only to fingerprint based matching, due to the nature of the reader included in the smart card; future development of smart card based biometric readers can add support for additional biometric features

For biometric passports, the most appropriate is the match-on-card technology.

The template-on-card technology has the disadvantage of allowing the personal biometric information to be retrieved from the smart card. In a closed environment, where all the components of the system can be thoroughly audited to prevent security leaks, that might not raise many issues, but considering the fact that passport identification represents a cross-border application, with multiple different implementations and tools, and many jurisdictions involved, it can raise numerous privacy issues each time the smart card is used.

The system-on-card technology might be the safest to use regarding the privacy, but is the least extendible, since it is intrinsically related to the fingerprint readers. It is also the least developed and market-ready technology of the three above. Cost is also prohibitive, and the current chip technologies that can be embedded in smart cards are either not fast enough to process pattern searches on the scanned fingerprint, or have high levels of power consumption that require complex solutions for embedding batteries into the smart card.

The match-on-card is extendible by adding support for other pattern matching than the fingerprint based one, and is safer to use, since it doesn't provide the stored biometric template to the readers, which is a very sensitive issue especially if the smart card is a contactless one, accessible without any physical contact between the smart card and the reader.

**Issues and Concerns**
The move to a biometric cross border identification system raises numerous concerns, regarding both the privacy of the citizens using the new identification system, and the implementation of such a system in different states.

Some of these concerns were raised by the European Data Protection Supervisor (EDPS) regarding the proposal for a regulation of the European Parliament for standards for security features and biometrics in passports and travel documents. Some of these concerns are presented in the following [4].

The European Commission failed to consult the EDPS regarding the proposal, showing little or no concern at all regarding the protection of individuals' rights and freedoms with regard to the processing of personal data. The Commission didn't conduct an impact assessment on the proposal, being therefore in unable to properly evaluate the necessity and proportionality of the proposal in relation to data protection issues, limiting the analysis to the cost triggered by new measures.

The use of biometrics provides advantages, but these benefits would be dependent on stringent safeguards being applied. A list of common obligations or requirements which need to be respected when biometric data are used in a system must be used to avoid that the passport holder is to carry the burden of system imperfections, such as misidentification or failure to enroll.

There should be exemptions from giving fingerprints based on the age of the person or his/her inability to give fingerprints. This is the case for children, for which the fingerprints are not of sufficient quality to allow on-to-one verification of identity. This also is the case of elderly people, with the accuracy and usability of fingerprints decreasing with age. Such persons shouldn't be discriminated by the proposed system.

Also, the one passport – one person concept, considering the exemptions above, fails to address the case of children below the age at which biometric data can be used travelling with their parents.

The identification documents used by the issuing agency to identify a person vary from state to state, including birth certificate, citizenship certificate, family book, parental authorization, driving license, utility bill. These documents are more likely to be subjected to forgery and counterfeiting, enjoying less security features. This can decrease the quality of data in passports for the entire system, and can lead to risks of identity theft. The biometric passport is only one link in the security chain starting with the identification documents used to obtain that passport, and therefore the entire system is just as safe as the weakest link, regardless of the enhanced security involved by the passport.

Also, as noted at the beginning of the article, several States have foreseen the implementation of a central database for storing the biometric data of the passport, which presents additional risks regarding the protection of personal data. The use of only decentralized storage, in the wireless chip of the passport, should be imposed by regulations.

Before going into effect such a proposal should address the issues of Failure to Enroll Rate (FER) for the enrollment process and False Rejection Rate (FRR) for the matching process in a uniform way for all the states.

**System Integration**
Identifying a person in a cross border system requires the use of well defined standards and protocols. The level of trust should very high for both the technologies used and the authorities issuing the identification documents. As stated above, the process of document issuing is basically an implementation of the public key infrastructure. Each country has a public agency that acts as a certificate authority. The identification document is the certificate that the agency has to authenticate.

The process starts by the citizen issuing a passport request. For this, he has to present full documentation to the authorities for national identification. A new identification card is created, which generates a new public/private keys pair. The person requesting the passport inputs the biometric data upon key generation, so that the cryptographic functions of the card will only be available upon authentication using the biometric elements (see figure 6).

The electronic certificate containing all the provided information, along with the public key and the picture, is sent to the national public authority, which validates it using the private key of the agency. This way any party interested can check the information provided by the user of the passport against the public key (usually enclosed in a certificate also) of the issuing agency.

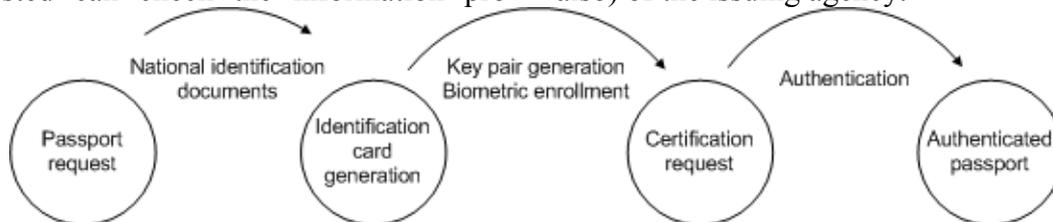

**Fig. 6.** Passport request

In order to check the identity of the passport owner, the owner will first present the passport to the border authority. Using the biometric readers, he will then give access to the internally stored certificate information (identification data and picture). The obtained certificate is validated against the certification authority (the issuing national agency). If the certificate is validated, a final sign request is sent to the card in order to verify that the smart card contains the private key paired with the public key from the certificate. If the response is valid, the process ends successfully (see figure 7).

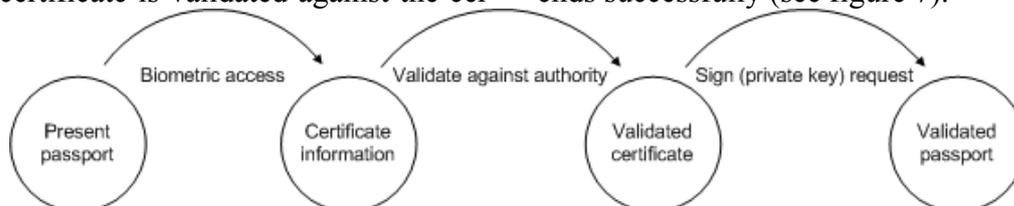

**Fig. 7.** Validating passport

The performed checks are not related only to the mathematical pairing of the keys and signatures. The national authority that issued the passport can, and should maintain a revocation list, against which the passport can be also automatically checked. The passports will also have expiration dates specified in the embodying certificate, which will be automatically checked by the system.

The only thing that the system should not do is store the biometric information anywhere. The only storage operation of biometric information should be performed during passport generation, and only on the smart card. This step should only be allowed once, when issuing the certificate. This can be ensured by the design of the smart card, and the fact that the information from a passport cannot be moved on a more permissive smart card, since the private key cannot be retrieved.

This way both the uniqueness of the passport information and the match between the smart card and its owner are proved with the high level of certainty required from the new electronic passport system.

**Conclusions**

The need for better identification methods is a valid and justified one. Technological advances allow the implementation of informational systems much more secure than a paper based authentication implementation, regardless of the security elements involved in the production of such paper based documents – one of the most used ways of falsifying paper documents is not generating new ones, but reusing existing ones.

The advances in the information technology, especially in the security and biometrics fields, allow national agencies to put behind all these concerns. But, with the switch to an electronic reimplementation of the existing system, special care has to be put into not raising new issues, most importantly regarding the privacy of those that should benefit from the system.

The solutions exist for avoiding these issues, and the society has proven their importance by taking action against any plan that ignored them. The approach presented in this article isn't the only possible one. It's intended only as a proof of concept that there is a way to solve all the current problems that led to the need to improve the system, without introducing new problems.

In avoiding introducing new problems resides the success of any change, and this one makes no exception from the rule.

**References**
[1] R. C. Bronk. *Innovation by Policy: A Study of the Electronic Passport.* Houston : The James A. Baker III Institute for Public Policy, Rice University, 2007.
[2] C. T. Clancy, Kiyavash, Negar and Lin, J. Dennis. "Secure Smartcard-Based Fingerprint Authentication," in *Proc. of Biometrics Methods and Applications, ACM SIGMM Workshop* (Berkley, California). New York, 2003.
[3] European Council. 2004. *Council Regulation (EC) No 2252/2004 of 13 December 2004 on standards for security features and biometrics in passports and travel documents issued by Member States.* 2004.
[4] European Data Protection Supervisor. 2008. *Opinion on the proposal for a Regulation of the European Parliament and of the Council amending Council Regulation (EC) No 2252/2004 on standards for security features and biometrics in passports and travel documents issued by Member States.* 2008.
[5] M. Hendry *Multi-application Smart Cards: Technology and Applications.* Cambridge : Cambridge University Press, 2007..
[6] Microsoft Corporation. 2005. X.509 Technical Supplement. *Microsoft Developer Network.* [Online] December 2005. [Cited: Feb. 1, 2009.] http://msdn.microsoft.com/en-us/library/aa480610.aspx.
[7] H. Nemati, *Information Security and Ethics: Concepts, Methodologies, Tools, and Applications.* Hershey: IGI Publishing, 2008.
[8] Ž. Požgaj and I. Đurinek. "Smart Card in Biometric Authentication," in *Proc. of. Information and Intelligent Systems*, Faculty of Organization and Informatics, University of Zagreb, Varaždin 2007.
[9] R. Rivest, A. Shamir, and L. Adleman, *Method for Obtaining Digital Signatures and Public-Key Cryptosystems.* Massachusetts Institute of Technology. New York : Association for Computing Machinery, 1978. Communications of the ACM.
[10] M. Toorani and Shirazi, A. A. Beheshti. *LPKI* "A Lightweight Public Key Infrastructure for the Mobile Environments," in Proc *of 11th IEEE Singapore International Conference on Communications Systems, 2008.*

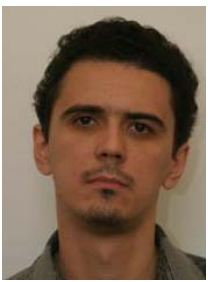
**Paul BĂLĂNOIU** has graduated the Faculty of Economic Cybernetics, Statistics and Informatics from the Bucharest Academy of Economic Studies in 2005. He is currently conducting doctoral research in Economic Informatics at the Academy of Economic Studies. He has a background in computer science, including numerous enterprise-grade projects ranging from Enterprise Resource Planning systems to document workflow management and quality of service monitoring. Other areas of interest are: information society, network security, electronic archives, document management, and judicial system.